\documentclass[entropy,article,accept,pdftex,10pt,a4paper]{mdpi}%
\usepackage{graphicx}
\usepackage{amsmath}
\usepackage{bbm}
\usepackage{amsfonts}
\usepackage{amssymb}
\usepackage{slashbox}%
\providecommand{\U}[1]{\protect\rule{.1in}{.1in}}
\newtheorem{theorem}{Theorem}
\newtheorem{proposition}[theorem]{Proposition}

\lastpage{25}
\doinum{10.3390/e19050202}
\pubvolume{19(5), 202}
\pubyear{2017}
\history{Submitted 12 March 2017 / Published: 2 May 2017 (version 2)}

\Title{Kinetics of interactions of matter, antimatter and radiation consistent with antisymmetric (CPT-invariant) thermodynamics.}

\Author{A. Y. Klimenko}

\address{%
The University of Queensland, QLD 4072, Australia}

\corres{E-mail: klimenko@mech.uq.edu.au}

\abstract{ 
This work investigates the influence of directional properties of decoherence on kinetics rate equations. The physical reality is understood as a chain of unitary and decoherence events. The former are quantum-deterministic, while the latter introduce uncertainty and increase entropy.  For interactions of matter and antimatter, two approaches are considered: symmetric decoherence, which corresponds to conventional  symmetric (CP-invariant) thermodynamics, and antisymmetric decoherence, which corresponds to antisymmetric (CPT-invariant) thermodynamics.  
Radiation, in its interactions with matter and antimatter, is shown to be decoherence-neutral. The symmetric and antisymmetric assumptions result in different interactions of radiation with matter and antimatter. The theoretical predictions for these differences are testable by comparing absorption (emission) of light by thermodynamic systems made of matter and antimatter.  Canonical typicality for quantum mixtures is briefly discussed in the Appendix.  
    }

\keyword{thermodynamic time, CPT-invariance, negative temperatures, kinetic equation, interaction of radiation with matter}

\begin{document}

%\maketitle

\section{Introduction}

The existence of the arrow of time (i.e. physical dissimilarity of the
directions of time) and its relation to the laws of the universe is one of the
most fundamental and still unresolved problems in modern physics
\cite{Boltzmann-book,PriceBook,PenroseBook,Zeh2007}. This dissimilarity is
likely to be "primed" by as-yet undiscovered small-scale physical processes,
which are likely to be conceptually related to interactions of quantum and
thermodynamic principles. These principles have been repeatedly discussed in
publications \cite{PenroseBook,Abe2011,QantumThermo2016}. One of the open
questions at the intersection of thermodynamics and quantum mechanics is that
traditional thermodynamics can be extended from matter to antimatter in two
possible mutually exclusive ways --- symmetric and antisymmetric --- and it is
not known which one of these extensions corresponds to the real world
\cite{KM2014}. The symmetric extension is CP-invariant and conventional, while
the antisymmetric extension is CPT-invariant and is of prime interest in the
present work, particularly in the context of kinetics rate equations for
antisymmetric thermodynamics. The symmetric and antisymmetric versions of
thermodynamics correspond to different directions of thermodynamic time and,
at the quantum level, to different predominant temporal directions of
decoherence, which can be either symmetric or antisymmetric with respect to
the duality of matter and antimatter \cite{SciRep2016}. The terminology based
on CP- and CPT-invariance was previously introduced for the two versions of
thermodynamics \cite{KM2014} following the discussion of presumed macroscopic
invariant properties of the universe\ \cite{Sakharov1967,PenroseBook}. Further
investigation of microscopic properties of quantum systems in the presence of
decoherence \cite{SciRep2016} determined that the expected thermodynamic
invariance is generally not linked to CP- and CPT-invariance of Hamiltonians
controlling unitary evolutions of quantum systems. To avoid confusion, in the
present work we refer to CP-invariant thermodynamics and kinetics as
"symmetric" and to CPT-invariant thermodynamics and kinetics as
"antisymmetric". The main feature of antisymmetric thermodynamics is the
existence of two temperatures of antimatter, intrinsic $\bar{T}$ and apparent
$T=-\bar{T}.$ The latter is revealed in interactions of antimatter with
matter. In symmetric thermodynamics both temperatures are the same $T=\bar{T}$.

The kinetics of mutual conversion of matter and antimatter (which practically
implies the existence of baryon number violations) has been investigated in
Ref.\cite{SciRep2016}, where two forms of Pauli master equation ---
conventional symmetric, corresponding to symmetric thermodynamics, and
non-conventional antisymmetric, corresponding to antisymmetric thermodynamics
--- have been derived. The derivation is based on the Pauli approach
\cite{Pauli1928}, which involves separation of quantum interactions into a
chain of reversible unitary evolution and irreversible decoherence events. The
present work extends this approach to a different problem: energy exchange
between matter and antimatter under conditions when mutual conversions of
matter and antimatter are not allowed.

The interaction of matter and antimatter with radiation is another principal
question considered here. The antisymmetric approach to thermodynamics
constrains these interactions by expecting temporal neutrality of radiation
(that is, intrinsically, neither decoherence nor its time inverse ---
recoherence --- can dominate the other for radiation interacting with matter
or antimatter). This work derives the rate equations for interactions of
radiation with matter and antimatter and demonstrates that only those assuming
that radiation is decoherence-neutral correspond to the real world.

The process of decoherence is a microscopic factor that may involve both
intrinsic \cite{Zurek2002, QTreview,Beretta2005,Stamp2012} and environmental
mechanisms
\cite{Stamp2012,Zurek1982,Joos2003,EnvDec2005,CT-G2006,CT-P2006,Yukalov2012}.
We generally presume the existence of both mechanisms and do not dwell on
physical causes of decoherence since the exact mechanisms of decoherence
remain unknown \cite{PenroseBook,Stamp2012}. It is also not known which
micro-objects may display thermodynamic behaviour. Conventional wisdom expects
that thermodynamic properties are associated only with macroscopic, not
microscopic, objects. This wisdom has been recently challenged by discovering
a thermodynamic (or at least thermodynamic-like) behaviour in high-energy
collisions of baryons resulting in emergence of quark-gluon plasma
\cite{Nature2007}.

We must note that antisymmetric interpretation of thermodynamics and kinetics
may seem unusual to many: causality is deeply embedded into our intuition and
often implicitly brings time-directional bias into thinking. Typically, this
happens when a problem is implied to possess initial and not final conditions
--- this interpretation corresponds well to our intuition. This point can be
illustrated by the following example: the stochastic trajectories $z(t)$
generated by Brownian motion are time symmetric; it is impossible to
distinguish $z(t)$ from $z(-t)$ for a given realisation. Yet the diffusion
equation for the probability distribution $P(z)$ is strongly time-directional.
The reason behind this is not temporal directionality of $z(t)$ but setting
initial (and not final) conditions for $z$, say, $z=0$ at $t=0$. It is not
wrong, of course, to use initial conditions --- this generally corresponds to
our predominant experience obtained in the real world --- but, in the context
of antisymmetric thermodynamics, this needs to be done explicitly with
understanding of the implications for possible violation of temporal symmetry.
The casual behaviour is to be purposely analysed and not implicitly presumed
\cite{PriceBook}. The need to suppress our causality-driven intuition is,
perhaps, aligned better with time-neutral approaches to quantum mechanics,
such as canonical typicality \cite{CT-G2006,CT-P2006} and time-symmetric
two-state quantum mechanics \cite{2S-QM1964, Wharton2007,2S-QM2008}.

The paper is organised as follows. Section 2 introduces a framework for
analysis of the influence of decoherence and recoherence on kinetics rate
equations. Kinetics of energy exchanges between matter and antimatter, which
is driven by symmetric or antisymmetric thermodynamics, is considered in
Section 3. Section 4 analyses the decoherence-related properties of
interactions of radiation with matter and these results are used in Section 5
to derive rate equations for interactions of radiation with antimatter. A
general discussion is presented in Section 6. Canonical typicality and
thermalisation are briefly discussed in the Appendix.

\section{Chain of unitary and decoherence events}

The effect of decoherence is profound and, as demonstrated for the Pauli
master equation, the presumed direction of decoherence ultimately controls the
thermodynamic direction of time \cite{SciRep2016}. The present work extends
these ideas to particle systems.

\begin{figure}[h]
\begin{center}
\includegraphics[width=10cm,page=1,trim=2cm 2cm 2cm 8cm, clip ]{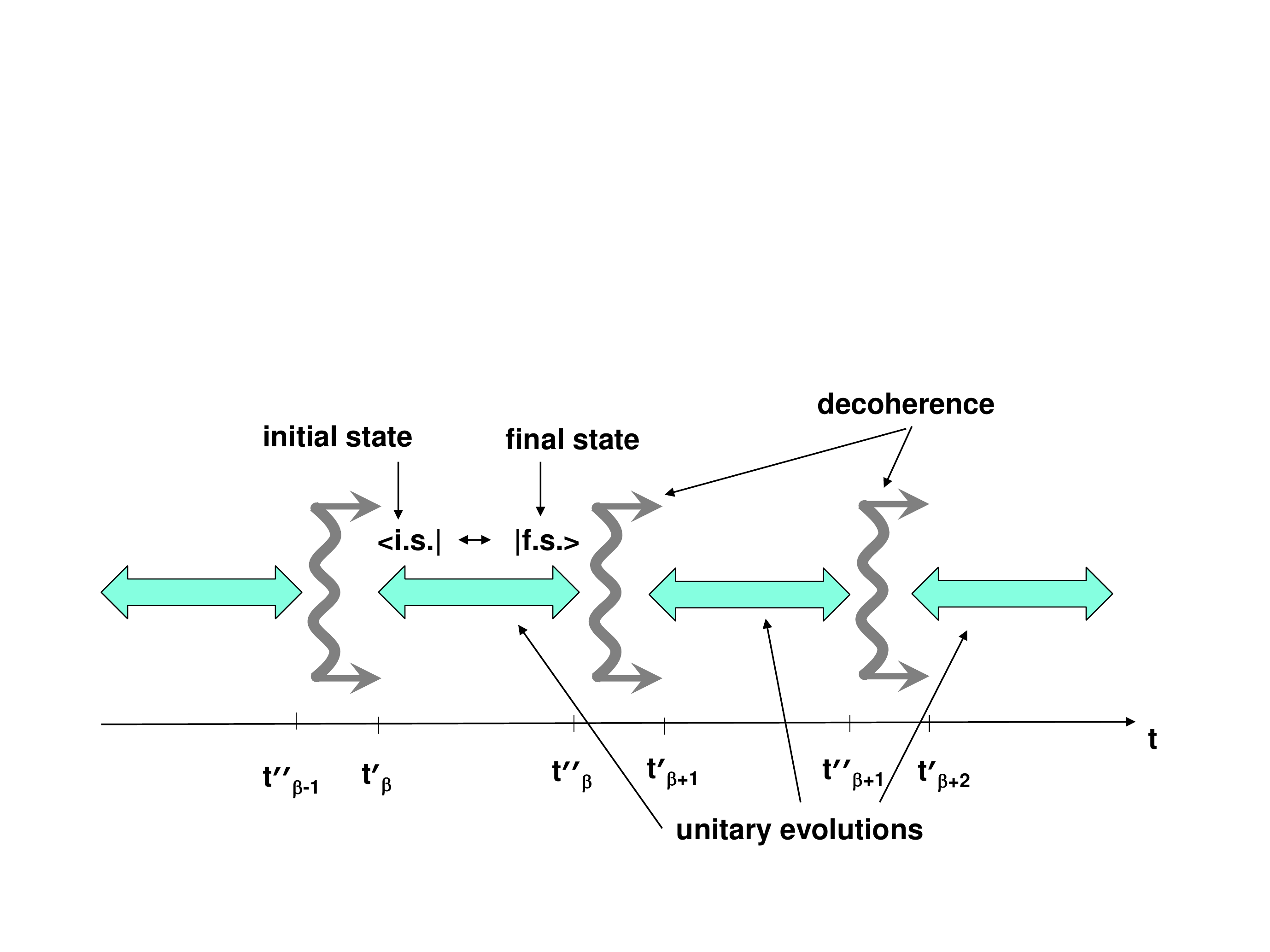}
\caption{Sequence of decoherence and unitary events in a reacting system.}
\label{fig1}
\end{center}
\end{figure}

The system considered here experiences two type of events: statistical, which
can increase the phase volume occupied by the system, and quantum mechanical,
which predict reversible (unitary) evolution of subsystems. Following the
previous work \cite{SciRep2016}, which is based on the approach originated by
Pauli \cite{Pauli1928}, unitary evolutions $\mathfrak{U}_{\beta}$ and
decoherence events $\mathfrak{D}_{\beta}$ are presumed to occur as a chain
$...,\mathfrak{U}_{\beta},\mathfrak{D}_{\beta},\mathfrak{U}_{\beta
+1},\mathfrak{D}_{\beta+1},...$, separated by the time moments $...,t_{\beta
}^{\prime},t_{\beta}^{\prime\prime},t_{\beta+1}^{\prime},t_{\beta+1}%
^{\prime\prime},...$ as illustrated in Figure \ref{fig1}. The events
$\mathfrak{U}_{\beta}$ are reversible and entropy-preserving, while randomness
and temporal directionality are introduced by events $\mathfrak{D}_{\beta}$,
which add uncertainty and, ultimately, increase entropy of the system. This
chain division of complex interactions is, of course, an idealisation of the
physical reality but it is needed to separate and distinguish the influences
of $\mathfrak{U}_{\beta}$ and $\mathfrak{D}_{\beta}$ and make the problem
tractable. Such idealisations, however, are common in quantum mechanics: for
example, separating unitary evolutions and non-unitary measurements. In
reality, decoherence is likely to be a combined effect of tiny decohering
violations of unitary mechanics combined with some unitary interactions
(intrinsic or environmental). In this work, we tend to bypass the complex
problem of the physical mechanism of decoherence and focus on its effect.

The question of quantum reality is complicated but, in the context of the
present work, the quantum events are connected to perceivable reality only
through the initial $\left\vert i,t_{\beta}^{\prime}\right\rangle $ and final
states $\left\vert f,t_{\beta}^{\prime\prime}\right\rangle $; everything else
in relation to the trajectories remains unknown and may be subject to
different physical interpretations (we can refer to two-state quantum
mechanics \cite{2S-QM1964, Wharton2007,2S-QM2008} as one of these
interpretational possibilities). Superposition of all these trajectories
contribute to the Feynman path integral which determines the statistical
weights of the events. Each unitary event has its statistical weight
$W_{i}^{f}$ defined by $W_{i}^{f}=\left\vert \hat{S}_{i}^{f}\right\vert ^{2}$
where $\hat{S}_{i}^{f}=\left\langle f|\mathbb{U}(\tau_{\beta})|i\right\rangle
$ is the scattering matrix, $\mathbb{U}(\tau_{\beta})$ reflects unitary
evolution over period $\tau_{\beta}=t_{\beta}^{\prime\prime}-t_{\beta}%
^{\prime}.$ For thermodynamic considerations, the \^{S}-matrix is assumed to
be close to the unit matrix: $\mathbf{\hat{S}}=\mathbf{I}+i\mathbf{h}$ and
$\left\Vert \mathbf{h}\right\Vert \ll\left\Vert \mathbf{I}\right\Vert $. Under
these conditions, $\mathbf{h}$ must be Hermitian as long as $\mathbf{\hat{S}}$
is unitary: $\mathbf{I}=\mathbf{\hat{S}}^{\dag}\mathbf{\hat{S}=I+}%
i(\mathbf{h}-\mathbf{h}^{\dag})+...$ The matrix $\mathbf{W}$ is then
symmetric
\begin{equation}
W_{i}^{f}=W_{f}^{i} \label{Wsym}%
\end{equation}
which is important for thermodynamic consistency \cite{SciRep2016}. The
smallness of $\mathbf{h}$ implies that the decoherence events are sufficiently
frequent but, as discussed below, not excessively frequent. Note that some of
the events $\left\vert i\right\rangle $ $\longrightarrow\left\vert
f\right\rangle $ might be physically impossible (say, a transition violating a
conservation law); in this case $W_{i}^{f}=W_{f}^{i}=0$.

Evaluation of the scattering matrix using quantum perturbation techniques have
been repeatedly discussed in publications in general
\cite{Pauli1928,Dyson1949,LL3} and specifically in application to interactions
of radiation and matter
\cite{Einstein1917,Dirac1927,Fermi1932,LL4,Andrews1989,Griff2005,Salam2010}.
In these theories, the system Hamiltonian $\mathbb{H}$ is represented by a sum
of the undisturbed component $\mathbb{H}_{0}$ and interaction component
$\mathbb{H}_{1}$ so that $\mathbb{H=H}_{0}\mathbb{+H}_{1}$ with different
corresponding characteristic times $\tau_{0}\sim1/\left\vert \mathbb{H}%
_{0}\right\vert \ll\tau_{1}\sim1/\left\vert \mathbb{H}_{1}\right\vert $
associated with these components. The characteristic time $\tau_{\beta}$
between the decoherence events is presumed to satisfy $\tau_{0}\ll\tau_{\beta
}\ll\tau_{1}$ (see ref.\cite{SciRep2016}): condition $\tau_{\beta}\ll\tau_{1}$
is required for thermodynamic consistency (\ref{Wsym}), while condition
$\tau_{0}\ll\tau_{\beta}$ enforces the conservation of energy in quantum
transitions and eliminates the quantum Zeno effect. When evaluating the
interaction Hamiltonian, atoms are commonly represented at the leading order
by electric dipoles since the wavelength of radiated light significantly
exceeds the typical size of the atoms.

Figure \ref{fig1} illustrates a conventional (symmetric) case when decoherence
occurs only in one temporal direction. The present work, however, allows for
another possibility, which represents decoherence occurring backward in time
--- this process is called here recoherence (although term "recoherence" can
have different interpretations elsewhere). In accordance with antisymmetric
thermodynamics, recoherence is expected to be dominant for antimatter, while
matter is dominated by decoherence. The symmetric version of thermodynamics
corresponds to the same predominant direction of decoherence for both matter
and antimatter.

The analysis of symmetric and antisymmetric master equations \cite{SciRep2016}
indicate that, for any selected segment of unitary evolution $\mathfrak{U}%
_{\beta},$ the conversion rates are determined by the amplitudes of decohered
components, whose presence depends on the directional properties of the
decohering events $\mathfrak{D}_{\beta}$. There is, however, another
constraint, which is linked to the presence of the recohered components. In
simple terms, the recohered components of a reaction do not affect the
probabilities and the rates of conversion as long as they can be present
without violating any physical conservation laws. Depending on temporal
direction of the kinetics rate equations, this constraint may or may not be
satisfied automatically. Practically, this constraint terminates the
conversions when the boundaries of physically possible values of the
parameters are reached (e.g. concentrations of the components must be
non-negative). The understanding offered by the analysis of Ref.
\cite{SciRep2016} can be summarised in the following proposition:

\begin{proposition}
\label{P1}In decohering systems, the conversion rate is proportional to the
probability magnitudes of the decohered components and, generally, is also
constrained by the presence of the required recohered components (although the
probability amplitudes of recohered components do not affect the conversion
rate). Decohered (recohered) components are understood as properties of
components immediately preceding or immediately following the segments of
unitary interaction, irrespective of the temporal direction of decoherence (recoherence).
\end{proposition}

There are, however, some features that distinguish present consideration from
the previous analysis \cite{SciRep2016} (where thermalisation was derived and
not presumed). As discussed in the Appendix, the system considered here is
assumed to be in a thermodynamic state, i.e. close to maximally mixed
conditions (although not necessarily exactly in maximally mixed conditions,
since this would imply achieving a thermodynamic equilibrium for an isolated
system). This assumption allows us to treat two-particle states as separable
(statistically independent), which is used in the further analysis.

\section{Kinetics and thermodynamics of indirect interactions of matter and
antimatter}

This section investigates kinetics and thermodynamics of relatively weak
interactions of matter and antimatter, when matter and antimatter are not in
direct contact and not allowed to annihilate. This, however, does not prohibit
a limited exchange of energy by virtual photons with energy $\Delta E$ as
shown in Figure \ref{fig3}. The matter and antimatter are represented by atoms
A and by antiatoms \={A} that correspondingly belong to a thermodynamic system
and antisystem (i.e. a system made of antimatter). Each of the atoms and
antiatoms can be in excited or ground state and each of these states is not
degenerate. The system under consideration is constrained by having only two
energy states, since these relatively simple systems provide the best
illustration for interactions of thermodynamics and kinetics. The excited
state may also be referred to as the "roof state" to emphasise that this state
corresponds to the highest possible energy in the system.

The numbers of atoms conserve the amounts of matter and antimatter
\begin{equation}
F_{\text{A}^{\ast}}+F_{\text{A}^{\circ}}=F_{\text{A}}=\operatorname{const}%
\text{ and }F_{\text{\={A}}^{\ast}}+F_{\text{\={A}}^{\circ}}=F_{\text{\={A}}%
}=\operatorname{const} \label{consm}%
\end{equation}
and the energy%
\begin{equation}
F_{\text{A}^{\circ}}+F_{\text{\={A}}^{\circ}}=F^{\circ}=\operatorname{const}%
\text{ and }F_{\text{A}^{\ast}}+F_{\text{\={A}}^{\ast}}=F^{\ast}%
=\operatorname{const} \label{conse}%
\end{equation}
The energy exchange reactions
\begin{equation}
1)\text{\ A}^{\ast}+\text{\={A}}^{\circ}\longrightarrow\text{A}^{\circ
}+\text{\={A}}^{\ast} \label{R1}%
\end{equation}%
\begin{equation}
2)\ \text{A}^{\circ}+\text{\={A}}^{\ast}\longrightarrow\text{A}^{\ast
}+\text{\={A}}^{\circ} \label{R2}%
\end{equation}
are shown in Figure \ref{fig3} and involve an energy transition through a
virtual photon (the first reaction is on left-hand side, and the second is on
right-hand side). The reaction rate constants are the same for forward and
reverse reactions due to thermodynamic consistency (\ref{Wsym}) (or due to
CP-invariance of quantum interaction Hamiltonians). The change in particle
numbers due to reactions over the intervals $[t_{\beta}^{\prime},t_{\beta
}^{\prime\prime}]$ is given by
\begin{equation}
\Delta F_{\text{A$^{\circ}$}}=F_{\text{A$^{\circ}$}}+\Delta F,\text{\ \ }%
\Delta F_{\text{A}^{\ast}}=F_{\text{A}^{\ast}}-\Delta F,\ \ \ \Delta
F_{\text{\={A}$^{\circ}$}}=F_{\text{\={A}$^{\circ}$}}-\Delta F,\text{\ \ }%
\Delta F_{\text{\={A}}^{\ast}}=F_{\text{\={A}}^{\ast}}+\Delta F
\label{ABBA_df}%
\end{equation}
where $\Delta F=\Delta F_{1}-\Delta F_{2}$ is the effect of both reactions
(\ref{R1}) and (\ref{R2}).

While the number of the energy states is limited to two, the atoms and
antiatoms can be found in $N$ secondary quantum states $J=1,...,N$, which do
not affect the energy but reflect uncertainty present in any thermodynamic
system. The value $N$ is presumed to be very large (compared to the number of
particles, which is also large) allowing us the use of classical statistics.
Hence, $f_{\text{X}}=F_{\text{X}}/N\ll1$ for any X\ =\ A$^{\circ},$ A$^{\ast
},$ \={A}$^{\circ}$ or \={A}$^{\ast}$.

The overall system states form a Hilbert space of very large dimension, which
can be estimated by $n_{0}\sim N^{F_{\text{A}}+F_{\text{\={A}}}}/(F_{\text{A}%
}!F_{\text{\={A}}}!)$, but such description would be excessively detailed and
is not needed. Energy exchange occurs only between atom/antiatom pairs ---
interactions between three and more particles are unlikely and can be
neglected. Hence, only two-particle states need to be considered, which
generally correspond to every one of the $F_{\text{A}}F_{\text{\={A}}}$
possible pairs distributed over the $N^{2}$ possible states $J_{\text{A}%
},J_{\text{\={A}}}=1,...,N$. The system is assumed to be close to a maximally
mixed state with respect to the secondary states implying that $F_{\text{A}%
}F_{\text{\={A}}}$ pairs are uniformly distributed between $N^{2}$ quantum
states (see Appendix). In this case, the probability of a selected
two-particle state being occupied by an atom A and an antiatom \={A} is given
by $P=f_{\text{A}}f_{\text{\={A}}}$. The energy states, however, are not
necessarily at equilibrium and the system can evolve.

The probability of reactions (\ref{R1}) and (\ref{R2}) is determined by
statistical weights $w=W_{\ast\circ}^{\circ\ast}=W_{\circ\ast}^{\ast\circ},$
where $W_{ab}^{ba}=\left\vert \hat{S}_{ab}^{ba}\right\vert ^{2}$, and $\hat
{S}_{ab}^{ba}=\left\langle \text{A}^{b}\text{+\={A}}^{a}|\mathbb{U}%
(\tau_{\beta})|\text{A}^{a}\text{+\={A}}^{b}\right\rangle $, while $a$ and $b$
denote "$\ast$" or "$\circ$". Many of the elements of matrix $\mathbf{W}$ are
zeros since they correspond to impossible events. For example, $W_{\ast\ast
}^{\circ\ast}=0$ corresponds to reaction A$^{\ast}+$\={A}$^{\ast
}\longrightarrow$A$^{\circ}+$\={A}$^{\ast}$ that violates conservation of
energy. Generally, $w$ can depend on secondary sates $J_{\text{A}}$ and
$J_{\text{\={A}}}$ or even imply a transition between secondary states
$\left\vert ab,J_{\text{A}}^{\prime}J_{\text{\={A}}}^{\prime},t=t^{\prime
}\right\rangle \longrightarrow\left\vert ba,J_{\text{A}}^{\prime\prime
}J_{\text{\={A}}}^{\prime\prime},t=t^{\prime\prime}\right\rangle $ --- this
does not affect the structure of the final equations since we are interested
only in following the energy states. Note that the majority of two-particle
states are not occupied since $F_{\text{X}}\ll N,$ that $w\ll1$ and that $w=0$
for the most of $\mathbf{J}=(J_{\text{A}}^{\prime},J_{\text{\={A}}}^{\prime
},J_{\text{A}}^{\prime\prime},J_{\text{\={A}}}^{\prime\prime})$; hence the
expected number of conversion events is relatively small $\Delta F\ll
F_{\text{X}}$.

The probability amplitudes for energy transfer between atoms by a virtual
photon can be evaluated for specific conditions of the interactions. Assuming
that two free atoms are located at the distance $r,$ the probability
amplitudes have been repeatedly evaluated by quantum perturbation techniques
\cite{Andrews1989,Salam2010}. Unlike real photons, the virtual photons are
allowed to violate the four-momentum expression for massless particles and,
perhaps, represent more a mathematical perturbation term than a real physical
object. The perturbation theories \cite{Andrews1989,Salam2010} use the dipole
approximation and determine that, as the distance $r$ increases, the virtual
photon becomes more and more "real" while the strength of interaction scales
as $r^{-6}$ in the near (radiationless) zone and as $r^{-2}$ in the far
(radiational) zone. In general, both of these regimes are consistent with the
consideration presented in this section. This consideration, however,
necessarily requires that the photon (purely virtual in the near zone or
quasi-real in the far zone) is in a jointly coherent state with the atoms,
while the transition between the moment of emission and the moment of
absorption depicted in Figure \ref{fig3} remains unitary and is not affected
by decoherence. This implies that the characteristic decoherence time is
assumed to be longer that the characteristic life time of the quasi-real
photons. Unlike the "real" photons considered in the following sections, the
virtual photons illustrated by Figure \ref{fig3} do not represent objects
independently controlled by the laws of statistical physics. The atoms and
antiatoms, however, decohere (or recohere) after interaction and thus become
true statistical objects, as considered in the rest of this section.

\subsection{Symmetric kinetics}

\begin{figure}[h]
\begin{center}
\includegraphics[width=7cm,page=2,trim=4cm 0cm 3.5cm 0cm, clip ]{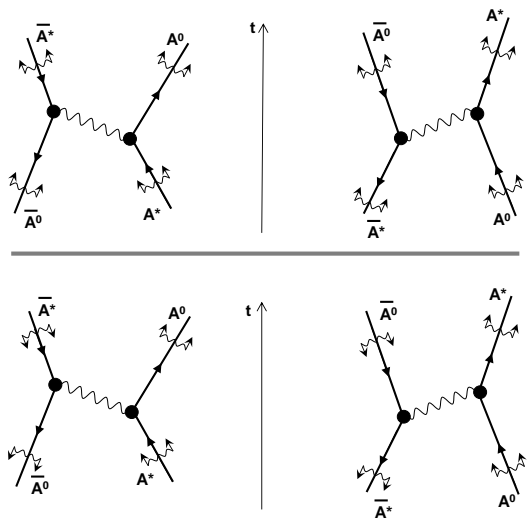}
\caption{Energy exchange between particle and antiparticle in a thermodynamic system with the same direction of decoherence: forward (left) and reverse (right) reactions. Top: symmetric (CP-invariant) decoherence; bottom: antisymmetric (CPT-invariant) decoherence. The braced arrows show the direction of decoherence before and after the interaction.}
\label{fig3}
\end{center}
\end{figure}

The case of symmetric decoherence, which corresponds to symmetric
thermodynamics, is shown in Figure \ref{fig3} (top). In this case, the
decohered components are represented by reactants and the recohered components
are represented by products. Hence, according to Proposition \ref{P1}, the
average number of convention events during the time interval $[t_{\beta
}^{\prime},t_{\beta}^{\prime\prime}]$ is given by
\begin{equation}
\Delta F_{1}=K_{1}f_{\text{A}^{\ast}}^{\prime}f_{\text{\={A}}^{\circ}}%
^{\prime}H\left(  f_{\text{A}^{\circ}}^{\prime\prime}\right)  H\left(
f_{\text{\={A}}^{\ast}}^{\prime\prime}\right)  \label{df1ff}%
\end{equation}%
\begin{equation}
\Delta F_{2}=K_{2}f_{\text{A}^{\circ}}^{\prime}f_{\text{\={A}}^{\ast}}%
^{\prime}H\left(  f_{\text{A}^{\ast}}^{\prime\prime}\right)  H\left(
f_{\text{\={A}}^{\circ}}^{\prime\prime}\right)  \label{df2ff}%
\end{equation}
for reactions (\ref{R1}) and (\ref{R2}) correspondingly. Here, where $H(F)=1$
when $F>0$ and $H(F)=0$ when $F\leq0$ is the Heaviside step function, the
"prime" superscript denotes parameters of the system at $t=t_{\beta}^{\prime}$
and "double prime" at $t=t_{\beta}^{\prime\prime}$. The reaction rate is
determined by the magnitudes of the decohered components ($f_{\text{A}^{\ast}%
}^{\prime}$ and $f_{\text{\={A}}^{\circ}}^{\prime}$\ for the first reaction),
which, as discussed in the Appendix, are deemed to be stochastically
independent, and constrained by the presence of recohered components
($f_{\text{A}^{\circ}}^{\prime\prime},f_{\text{\={A}}^{\ast}}^{\prime\prime
}>0$ for the first reaction). The reaction constants are given by
\begin{equation}
K_{1}=\sum_{\mathbf{J}}W_{\ast\circ}^{\circ\ast}(\mathbf{J})=\sum_{\mathbf{J}%
}W_{\circ\ast}^{\ast\circ}(\mathbf{J})=K_{2} \label{kkk}%
\end{equation}
irrespective whether the reactions affect the secondary states or not.

The difference between particle numbers before and after the reaction can
often be neglected, that is $F_{\text{X}}^{\prime\prime}\approx F_{\text{X}%
}^{\prime}\approx F_{\text{X}}$ and $f_{\text{X}}^{\prime\prime}\approx
f_{\text{X}}^{\prime}\approx f_{\text{X}}$ for all X since all $F_{\text{X}}$
are much larger then $1.$ Each reaction occurrence can change the numbers of
particles only by $1$. Given a reaction occurring in a particular box, the
overall change in particle numbers at a given time step is relatively low
$1+\left\vert \Delta F\right\vert \ll F_{\text{X}}$. This leads us to
possibility of converting equations (\ref{ABBA_df}), (\ref{df1ff}) and
(\ref{df2ff}) into an ordinary differential equation (ODE), which is often
called the reaction kinetics equation. While, generally, this conversion may
depend on whether the reaction rate equation is to be solved forward in time
(with initial conditions) or backward in time (with final conditions), the
forward-time and backward-time ODEs are equivalent in this case. With $\Delta
F=\Delta F_{1}-\Delta F_{2}$ in (\ref{ABBA_df}),\ the reaction rate equation
take the form
\begin{align}
-\frac{df_{\text{A}^{\ast}}}{dt}  &  =-\frac{df_{\text{\={A}}^{\circ}}}%
{dt}=\frac{df_{\text{A}^{\circ}}}{dt}=\frac{df_{\text{\={A}}^{\ast}}}%
{dt}\nonumber\\
&  =k_{\tau}\left(  f_{\text{A}^{\ast}}f_{\text{\={A}}^{\circ}}H\left(
f_{\text{A}^{\circ}}\right)  H\left(  f_{\text{\={A}}^{\ast}}\right)
-f_{\text{A}^{\circ}}f_{\text{\={A}}^{\ast}}H\left(  f_{\text{A}^{\ast}%
}\right)  H\left(  f_{\text{\={A}}^{\circ}}\right)  \right) \label{sk}\\
&  \underset{t\uparrow}{=}k_{\tau}\left(  f_{\text{A}^{\ast}}f_{\text{\={A}}%
^{\circ}}-f_{\text{A}^{\circ}}f_{\text{\={A}}^{\ast}}\right) \nonumber
\end{align}
where $k_{\tau}=K_{1}/\tau=K_{2}/\tau,$ $\tau=t_{\beta+1}-t_{\beta}$. For the
forward-time equation, the Heaviside functions can be omitted since all
populations $F$ remain non-negative as long as the initial conditions are also
non-negative. The Heaviside functions are nevertheless needed if equation
(\ref{sk}) is to be solved backward in time. Equation (\ref{sk}) is
CP-invariant and is not affected by swapping A and \={A}.

Note that the Gibbs distribution with any temperature $T$
\begin{equation}
\frac{f_{\text{A}^{\ast}}^{(e)}}{f_{\text{A}^{\circ}}^{(e)}}=\frac
{f_{\text{\={A}}^{\ast}}^{(e)}}{f_{\text{\={A}}^{\circ}}^{(e)}}=\exp\left(
-\frac{\Delta E}{T}\right)  \label{sT}%
\end{equation}
represents an equilibrium ($d/dt=0$) solution for equation (\ref{sk}). The
superscript "$(e)$" is used to indicate values related to equilibrium. The
thermodynamic ground state $(f_{\text{A}^{\ast}}^{(e)}=0$) corresponds to the
coldest possible temperature $T=+0$ and the thermodynamic roof state
$(f_{\text{A}^{\circ}}^{(e)}=0$) corresponds to the hottest possible
temperature $T=-0$. Note that, in systems with a finite number of energy
levels, the temperatures can be negative and that negative temperatures are
hotter than positive temperatures (\cite{Ramsey1956,LL5,K-OTJ2012}).

The conservation laws (\ref{consm}) and (\ref{conse}) in conjunction with
equilibrium condition (\ref{sT}) uniquely determine the equilibrium state by
\begin{equation}
\frac{f_{\text{A}^{\circ}}^{(e)}}{f_{\text{A}}}=\frac{f_{\text{\={A}}^{\circ}%
}^{(e)}}{f_{\text{\={A}}}}=\frac{f^{\circ}}{f_{\text{A}}+f_{\text{\={A}}}%
}\text{ and }\frac{f_{\text{A}^{\ast}}^{(e)}}{f_{\text{A}}}=\frac
{f_{\text{\={A}}^{\ast}}^{(e)}}{f_{\text{\={A}}}}=\frac{f^{\ast}}{f_{\text{A}%
}+f_{\text{\={A}}}} \label{se}%
\end{equation}
\ In this case $T$ is not fixed and is determined by (\ref{sT}).

With the use of conservation laws (\ref{consm}) and (\ref{conse}), the
symmetric kinetics rate equation specified by (\ref{sk}) can be rewritten in
the form
\begin{equation}
\frac{df_{\text{A}^{\circ}}}{dt}=k_{\tau}\left(  f_{\text{A}}+f_{\text{\={A}}%
}\right)  \left(  f_{\text{A}^{\circ}}^{(e)}-f_{\text{A}^{\circ}}\right)
\label{sk2}%
\end{equation}
It is clear that the system always converges to the equilibrium and that this
equilibrium is stable.

\subsection{Antisymmetric kinetics}

The case shown at the bottom of Figure \ref{fig3} corresponds to antisymmetric
decoherence and thermodynamics. Note that the model under consideration is
invariant with respect to swapping A and \={A} and changing $t$ to $-t$.\ As
the particles proceed from the past to the future, the atoms A are subject to
decoherence events (which impose probability constraints after decoherence),
while the antiatoms \={A} are subject to the recoherence events (which impose
the probability constraints before recoherence). According to Proposition
\ref{P1}, this leads to the following expressions for the average number of
(\ref{R1}) and (\ref{R2}) reaction events
\begin{equation}
\Delta F_{1}=K_{1}f_{\text{A}^{\ast}}^{\prime}f_{\text{\={A}}^{\ast}}%
^{\prime\prime}H\left(  f_{\text{A}^{\circ}}^{\prime\prime}\right)  H\left(
f_{\text{\={A}}^{\circ}}^{\prime}\right)  \label{df1ffa}%
\end{equation}%
\begin{equation}
\Delta F_{2}=K_{2}f_{\text{A}^{\circ}}^{\prime}f_{\text{\={A}}^{\circ}%
}^{\prime\prime}H\left(  f_{\text{A}^{\ast}}^{\prime\prime}\right)  H\left(
f_{\text{\={A}}^{\ast}}^{\prime}\right)  \label{df2ffa}%
\end{equation}
where $K_{1}=K_{2}=K$ are still defined by (\ref{kkk}). The reaction rate of
the first reaction is determined by decohered components $f_{\text{A}^{\ast}%
}^{\prime}$ and $f_{\text{\={A}}^{\ast}}^{\prime\prime}$ and constrained by
the presence of recohered components $f_{\text{A}^{\circ}}^{\prime\prime
},f_{\text{\={A}}^{\circ}}^{\prime}>0$.

As previously, the conversion of these relations into ODE, assuming
$F_{\text{X}}\gg1$ for all X and $k_{\tau}=K_{1}/\tau=K_{2}/\tau$, results in
the following reaction rate equation
\begin{align}
-\frac{df_{\text{A}^{\ast}}}{dt}  &  =-\frac{df_{\text{\={A}}^{\circ}}}%
{dt}=\frac{df_{\text{A}^{\circ}}}{dt}=\frac{df_{\text{\={A}}^{\ast}}}%
{dt}\nonumber\\
&  =k_{\tau}\left(  f_{\text{A}^{\ast}}f_{\text{\={A}}^{\ast}}H\left(
f_{\text{A}^{\circ}}\right)  H\left(  f_{\text{\={A}}^{\circ}}\right)
-f_{\text{A}^{\circ}}f_{\text{\={A}}^{\circ}}H\left(  f_{\text{A}^{\ast}%
}\right)  H\left(  f_{\text{\={A}}^{\ast}}\right)  \right) \label{sak}\\
&  \underset{f_{\text{X}}>0}{=}k_{\tau}\left(  f_{\text{A}^{\ast}%
}f_{\text{\={A}}^{\ast}}-f_{\text{A}^{\circ}}f_{\text{\={A}}^{\circ}}\right)
\nonumber
\end{align}
Here, we neglect the difference between $f^{\prime}$ and $f^{\prime\prime}$
since the change brought by each the reaction event is relatively small. The
Heaviside functions can be omitted for simplicity but the reactions need to be
terminated if any of the components is exhausted. The derived equation is
CPT-invariant: swapping A and \={A} and substituting $-t$ for $t$ does not
change equation (\ref{sak}).

Note that the Gibbs distribution with two intrinsic temperatures: $T$ for
matter and $\bar{T}$ for antimatter%
\begin{equation}
\frac{f_{\text{A}^{\ast}}^{(e)}}{f_{\text{A}^{\circ}}^{(e)}}=\exp\left(
-\frac{\Delta E}{T}\right)  ,\ \ \frac{f_{\text{\={A}}^{\ast}}^{(e)}%
}{f_{\text{\={A}}^{\circ}}^{(e)}}=\exp\left(  -\frac{\Delta E}{\bar{T}%
}\right)  ,\ \ T=-\bar{T} \label{saT}%
\end{equation}
corresponds to the equilibrium state ($d/dt=0$) of equation (\ref{sak}). The
conservation laws specified by (\ref{consm}) and (\ref{conse}) require that
equilibrium states are compliant with
\begin{equation}
\frac{f_{\text{A}^{\circ}}^{(e)}}{f_{\text{A}}}=\frac{f_{\text{\={A}}^{\ast}%
}^{(e)}}{f_{\text{\={A}}}}=\frac{f_{\text{\={A}}}-f^{\circ}}{f_{\text{\={A}}%
}-f_{\text{A}}}\text{ and }\frac{f_{\text{A}^{\ast}}^{(e)}}{f_{\text{A}}%
}=\frac{f_{\text{\={A}}^{\circ}}^{(e)}}{f_{\text{\={A}}}}=\frac{f^{\circ
}-f_{\text{A}}}{f_{\text{\={A}}}-f_{\text{A}}} \label{saee}%
\end{equation}
Note that physical equilibrium states exists only when both $f^{\circ}$ and
$f^{\ast}$ have values between $f_{\text{\={A}}}$ and $f_{\text{A}}$. The
conservation laws (\ref{consm}) and (\ref{conse}) transform (\ref{sak}) into
\begin{equation}
\frac{df_{\text{A}^{\circ}}}{dt}=k_{\tau}\left(  f_{\text{\={A}}}-f_{\text{A}%
}\right)  \left(  f_{\text{A}^{\circ}}^{(e)}-f_{\text{A}^{\circ}}\right)
\label{sak2}%
\end{equation}
where $f_{\text{A}^{\circ}}^{(e)}$ is formally defined by (\ref{saee})
irrespective of existence of a physical equilibrium state. The equation is
stable only if $f_{\text{\={A}}}>f_{\text{A}}$.\ If a physical equilibrium
does not exist or exists but is unstable, evolution of equation (\ref{sak2})
is terminated when at least one of the population numbers $f_{\text{A}^{\ast}%
}$, $f_{\text{A}^{\circ}}$, $f_{\text{\={A}}^{\ast}}$ or $f_{\text{\={A}}%
^{\circ}}$ becomes zero.

\subsection{Properties of antisymmetric kinetics}

The properties of the antisymmetric kinetics are examined below for three
characteristic cases $f_{\text{\={A}}}=f_{\text{A}}$, $f_{\text{\={A}}}\gg
f_{\text{A}}$ and $f_{\text{\={A}}}\ll f_{\text{A}},$ which are nominally
labelled "early universe", "travelling to antiworld" and "experiment with
antimatter". Practically, only the last case may be related to realistic
experiments or observations in present conditions.

\begin{itemize}
\item \textbf{Early universe:} $f_{\text{\={A}}}=f_{\text{A}}$. Equilibrium
under conditions of having the same amounts of matter and antimatter, which is
specified by $f_{\text{A}^{\circ}}=$ $f_{\text{\={A}}^{\ast}}$ and
$f_{\text{A}^{\ast}}=$ $f_{\text{\={A}}^{\circ}}$, is neutral and can be
achieved at different temperatures $T=-\bar{T}$ but is subject to the
additional condition $f_{\text{\={A}}^{\ast}}+f_{\text{A}^{\ast}}=$
$f_{\text{\={A}}^{\circ}}+$ $f_{\text{A}^{\circ}}$, which must be satisfied in
compliance with the solution $f_{\text{A}^{\circ}}=$ $f_{\text{\={A}}^{\ast}%
}=$ $f_{\text{A}^{\ast}}=$ $f_{\text{\={A}}^{\circ}}$ that has infinite
temperature $1/T=1/\bar{T}=0$.

\item \textbf{Travelling to antiworld:} $f_{\text{\={A}}}\gg f_{\text{A}}$. A
matter traveller of a small mass travels to an antiworld populated by large
amounts of antimatter (or the traveller is a fictional Time Lord and somehow
manages to turn his world line back in our world time). This case has a stable
thermodynamic equilibrium. Assuming that the intrinsic temperature of
antiworld is positive $\bar{T}>0$ this equilibrium can be achieved only at
negative temperatures $T=-\bar{T}$ of the traveller. Practically this means
that the traveller would be burned.

\item \textbf{Experiment with antimatter:} $f_{\text{\={A}}}\ll f_{\text{A}}$.
In this case equilibrium between matter and antimatter is unstable and
practically impossible. Depending on initial conditions, the antimatter object
would fall into the intrinsic ground state (apparent roof state) or, possibly
but much less likely, into the intrinsic roof state (apparent ground state).
\end{itemize}

These conclusions are based on kinetics rate equations but are in general
agreement with previous analysis of interactions matter and antimatter based
on thermodynamics. Note that the apparent heat capacity of antimatter is
negative in antisymmetric thermodynamics \cite{KM2014}. A brief summary of
thermodynamic behaviour for negative temperatures and negative heat capacities
can be found in ref. \cite{K-OTJ2012} and many other publications. It must be
mentioned that unlike in previous works, matter/antimatter annihilations and
baryon number violations are not allowed here, while matter and antimatter
interact only by exchange of radiation. The main results of the antisymmetric
treatment of the problem are summarised by this proposition:

\begin{proposition}
The antisymmetric (CPT-invariant) extension of thermodynamics and kinetics
from matter to antimatter corresponds to decoherence being dominant for matter
and recoherence being dominant for antimatter. Under these conditions,
thermodynamic equilibrium between matter and antimatter is theoretically
possible but practically unlikely since this requires negative intrinsic
temperatures. In most cases, a thermodynamic antisystem should fall into one
of its extreme states, most likely the intrinsic ground state.
\end{proposition}

\subsection{H-theorems for symmetric and antisymmetric kinetics}

The intrinsic entropy for each component is defined conventionally
\begin{equation}
S_{\text{X}}=-F_{\text{X}^{\circ}}\ln\left(  f_{\text{X}^{\circ}}\right)
-F_{\text{X}^{\ast}}\ln\left(  f_{\text{X}^{\ast}}\right)
,\ \ \ \ \text{X=A,\={A}} \label{Sx}%
\end{equation}
The kinetics considered in this section are compliant with the H-theorems as
formulated in the following proposition

\begin{proposition}
The symmetric $S_{s}=S_{\operatorname{A}}+S_{\operatorname{\bar{A}}}$ and
antisymmetric $S_{a}=S_{\operatorname{A}}-S_{\operatorname{\bar{A}}}$
entropies are increased forward in time $t$ by symmetric and antisymmetric
kinetics correspondingly until the system evolution reaches its equilibrium or
is terminated by physical constraints.
\end{proposition}

The entropy-increasing properties of conventional symmetric kinetics are
well-known --- the treatment of the antisymmetric case is shown below. We use
the indicator $\Theta_{\text{X}}$ that is defined $\Theta_{\text{X}}=1$ for
X=A$^{\circ},$A$^{\ast}$ and $\Theta_{\text{X}}=-1$ for X=\={A}$^{\circ}%
,$\={A}$^{\ast}$ and conclude:%
\begin{equation}
\frac{dS_{a}}{dt}=-\sum_{\text{X}}\Theta_{\text{X}}\frac{dF_{\text{X}}}%
{dt}\left(  \ln\left(  f_{\text{X}}\right)  +1\right)  =k_{\tau}\left(
f_{\text{A}^{\ast}}f_{\text{\={A}}^{\ast}}-f_{\text{A}^{\circ}}f_{\text{\={A}}%
^{\circ}}\right)  \ln\left(  \frac{f_{\text{A}^{\ast}}f_{\text{\={A}}^{\ast}}%
}{f_{\text{A}^{\circ}}f_{\text{\={A}}^{\circ}}}\right)  \geq0
\end{equation}
for any $f_{\text{A}^{\ast}},$ $f_{\text{\={A}}^{\ast}},$ $f_{\text{A}^{\circ
}}$ and $f_{\text{\={A}}^{\circ}}$.

\section{Interactions of radiation and matter.\label{Srm}}

In this section we consider interactions of radiation with matter, assuming
different decohering properties of radiation. Interactions of a single quantum
state populated by $q$ photons (which are bosons and do not restrict $q$) with
$F_{\text{A}}$ matter atoms specified by 1) absorption and 2) emission
reaction equations
\begin{equation}
\text{1) A}^{\circ}+\nu\longrightarrow\text{A}^{\ast}\text{ \ \ and \ \ 2)
A}^{\ast}\longrightarrow\text{A}^{\circ}+\nu\label{AnuA}%
\end{equation}
The probability amplitudes of absorption and emission have been extensively
studied in the literature \cite{Dirac1927,Fermi1932,LL4,Griff2005} and are
consistent with the emission/absorption probability amplitudes evaluated for
quasi-real photons \cite{Salam2010}. Unlike virtual and quasi-real photons of
the previous section, the photons and atoms considered here can decohere (or
recohere) and, thus, are both subject to the laws of statistical physics. Note
that photon populations $q$ can be both small and large, depending on
conditions. Assuming atom population per secondary state $f_{\text{A}^{a}%
}=F_{\text{A}^{a}}/N$ is at equilibrium, the canonical distribution is linked
to the temperature by
\begin{equation}
\frac{f_{\text{A}^{\ast}}}{f_{\text{A}^{\circ}}}=\exp\left(  -\frac{\Delta
E}{T}\right)  \label{Gibbs1}%
\end{equation}

\subsection{Radiation with prevailing decoherence}

\begin{figure}[h]
\begin{center}
\includegraphics[width=7cm,page=3,trim=6cm 2cm 6cm 1cm, clip ]{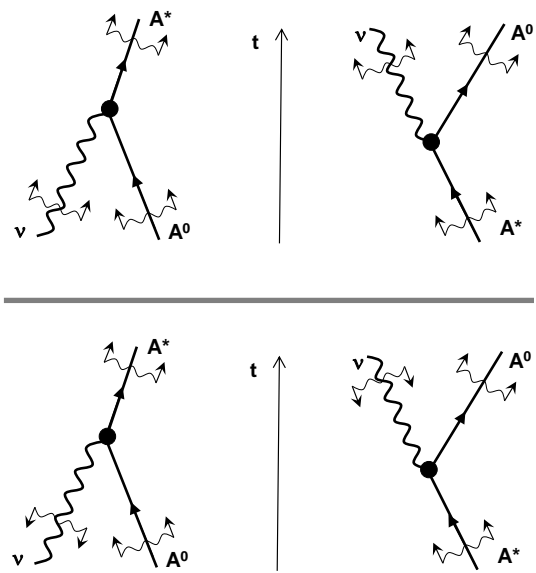}
\caption{Interactions of radiation with matter: absroption (left) and emission (right). Top: radiation with predominant decoherence; bottom: radiation with predominent recoherence.} 
\label{fig4}
\end{center}
\end{figure}

In this subsection we assume that radiation has the same decohering properties
as matter, which is illustrated in Figure \ref{fig4} (top). The single and
double primes are used to relate the values before and after each reaction:
there are $q^{\prime}$ photons before and $q^{\prime\prime}$ after the
reaction, as well as $F_{\text{A}^{\circ}}^{\prime}$ and $F_{\text{A}^{\circ}%
}^{\prime\prime}$ atoms before and after reactions in the ground state and
$F_{\text{A}^{\ast}}^{\prime}$ and $F_{\text{A}^{\ast}}^{\prime\prime}$ atoms
before and after reaction in the excited state. Note that $q^{\prime\prime
}=q^{\prime}-1$ for the absorption reaction and $q^{\prime\prime}=q^{\prime
}+1$ for the emission reaction. The absorption reaction is considered first.
As shown in Figure \ref{fig4} (top), all particles (atoms and photons)
decohere before and after the reactions. According to Proposition 1, the
number of reactions per reaction time $\tau$ is given by
\begin{equation}
\Delta q_{1}=Kf_{\text{A}^{\circ}}^{\prime}q^{\prime}H\left(  f_{\text{A}%
^{\ast}}^{\prime\prime}\right)  \text{ \ and\ }\Delta q_{2}=Kf_{\text{A}%
^{\ast}}^{\prime}H\left(  f_{\text{A}^{\circ}}^{\prime\prime}\right)  H\left(
q^{\prime\prime}\right)
\end{equation}
for the absorption and emission reaction correspondingly with $K$ being the
reaction constant and the overall effect on photon population given by $\Delta
q=\Delta q_{2}-\Delta q_{1}$.\ Since $q$ can be small, the differences between
$q^{\prime}$ and $q^{\prime\prime}$ cannot generally be neglected, although
assuming $f_{\text{A}^{a}}^{\prime\prime}=F_{\text{A}^{a}}^{\prime\prime
}/N\approx F_{\text{A}^{a}}^{\prime}/N=f_{\text{A}^{a}}^{\prime}$ is still
acceptable since $F_{\text{A}^{a}}\gg1$. Conversion of these relations into
ODEs is similar to the previous considerations but there is one point that
needs clarification. We seek to obtain a forward-time kinetics rate equation
as this equation is useful for us, since in the real world we set the initial
(and not final) conditions. This means that we can specify $q^{\prime}$ but
not $q^{\prime\prime}$, and must identify the current value of the population
$q$ with $q^{\prime}$ and put $q^{\prime\prime}=q+1$\ for the emission
reaction. This results in
\begin{equation}
\frac{dq}{dt}=kf_{\text{A}^{\ast}}H\left(  f_{\text{A}^{\circ}}\right)
H\left(  q+1\right)  -kqf_{\text{A}^{\circ}}H\left(  f_{\text{A}^{\ast}%
}\right)
\end{equation}
where $k=K/\tau$. The Heaviside constraints can be omitted. For example,
$H\left(  q+1\right)  =1$ since $q+1$ is always positive. This equation is not
consistent with the Einstein theory of radiation \cite{Einstein1917} and does
not reproduce the Bose-Einstein statistics. Indeed assuming Gibbs distribution
for atoms (\ref{Gibbs1}) we obtain for the equilibrium population%
\begin{equation}
q^{(e)}=\frac{f_{\text{A}^{\ast}}}{f_{\text{A}^{\circ}}}=\exp\left(
-\frac{\Delta E}{T}\right)
\end{equation}
that is inconsistent with the expected physical behaviour of radiation since
$q^{(e)}\rightarrow1$ as $T\rightarrow\infty$.

\subsection{Radiation with prevailing recoherence}

In this subsection, radiation is presumed to predominantly recohere, while
matter is still conventionally dominated by decoherence as shown in Figure
\ref{fig4} (bottom). The number of reactions per reaction time $\tau$ is given
by
\begin{equation}
\Delta q_{1}=Kf_{\text{A}^{\circ}}^{\prime}H\left(  f_{\text{A}^{\ast}%
}^{\prime\prime}\right)  H\left(  q^{\prime}\right)  \text{ and }\Delta
q_{2}=Kf_{\text{A}^{\ast}}^{\prime}q^{\prime\prime}H\left(  f_{\text{A}%
^{\circ}}^{\prime\prime}\right)
\end{equation}
This equation is converted into ODE by assuming, as previously, that
$q=q^{\prime},$ while taking into account that $q^{\prime\prime}=q^{\prime}+1$
in the emission reaction.\ The equation takes the form
\begin{equation}
\frac{dq}{dt}=k(q+1)f_{\text{A}^{\ast}}H\left(  f_{\text{A}^{\circ}}\right)
-kf_{\text{A}^{\circ}}H\left(  f_{\text{A}^{\ast}}\right)  H\left(  q\right)
\end{equation}
The constraint $H\left(  q\right)  $ in the last term prevents $q$ from
becoming negative, while the other Heaviside functions in this equation can be
omitted. With distribution (\ref{Gibbs1}) specified, the equilibrium
distribution for this equation is given by
\begin{equation}
q^{(e)}=\frac{f_{\text{A}^{\circ}}}{f_{\text{A}^{\ast}}}-1=\exp\left(
\frac{\Delta E}{T}\right)  -1
\end{equation}
which is obviously incorrect since $q^{(e)}\rightarrow0$ as $T\rightarrow
\infty$.

\subsection{Decoherence-neutral radiation.}

\begin{figure}[h]
\begin{center}
\includegraphics[width=7cm,page=4,trim=6cm 2cm 6cm 1cm, clip ]{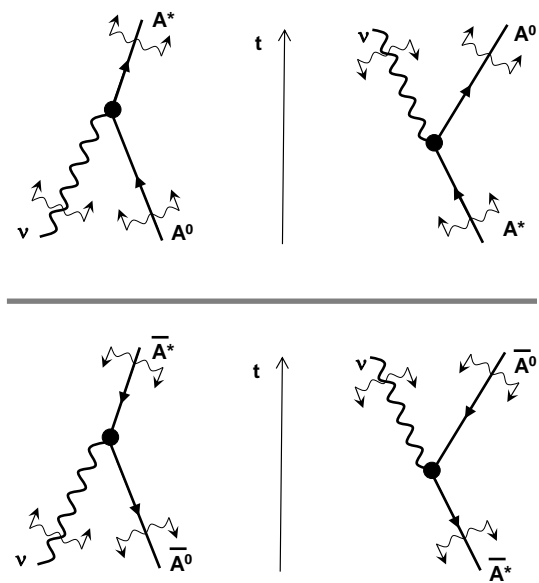}
\caption{Interactions of decoherence-neutral radiation with matter (top) and antimatter (bottom): absroption (left) and emission (right).} 
\label{fig5}
\end{center}
\end{figure}

The remaining option for radiation is\ to be decoherence-neutral, implying
that neither decoherence nor recoherence can dominate unconditionally.
Statistic nature of thermodynamic interactions, however, implies that
interactions of photons and matter must involve some degree of decoherence.
Decoherence-neutral interaction of photon with matter is depicted in Figure
\ref{fig5} (top). The atom decoheres both before and after emission or
absorption, while the photon decoheres before absorption and recoheres after
emission. The decohered states of the photon are caused not by intrinsic
mechanisms pertaining to the electromagnetic fields but by interactions with
matter. Under these conditions decoherence/recoherence of the photon in Figure
\ref{fig6} is effectively time-symmetric.

A possible physical scenario for decoherence-neutral radiation is that,
initially, the $q$ photons are in a coherent Fock state but they decohere due
to weak decohering interactions with atoms (or antiatoms). The decohered
photons are engaged into stronger energy interactions with atoms, which can
lead, although with a small probability, to absorption or emission of a
photon. When all interactions with matter are completed, the remaining photons
recohere back into a Fock state.

Absorption and emission of photons are now considered. There are $q^{\prime}$
photons before the reaction and $q^{\prime\prime}$ after the reaction.
According to the analysis of the previous sections, the number of reactions
per reaction time $\tau$ is given by
\begin{equation}
\Delta q_{1}=K\ f_{\text{A}^{\circ}}^{\prime}q^{\prime}H\left(  f_{\text{A}%
^{\ast}}^{\prime\prime}\right)  \label{rm1}%
\end{equation}
for the absorption reaction, where $F_{\text{A}^{\ast}}^{\prime\prime}\approx
F_{\text{A}^{\ast}}^{\prime}$ and $q^{\prime}=q^{\prime\prime}+1\neq
q^{\prime\prime}$ and
\begin{equation}
\Delta q_{2}=K\ f_{\text{A}^{\ast}}^{\prime}q^{\prime\prime}H\left(
f_{\text{A}^{\circ}}^{\prime\prime}\right)  \label{rm2}%
\end{equation}
for the emission reaction, where $F_{\text{A}^{\circ}}^{\prime\prime}\approx
F_{\text{A}^{\circ}}^{\prime}$ and $q^{\prime\prime}=q^{\prime}+1\neq
q^{\prime}$. Since photon population $q$ can be small, the differences between
$q^{\prime}$ and $q^{\prime\prime}$ cannot generally be neglected. Conversion
of these relations into ODE is similar to the previous considerations. As we
seek to obtain a forward-time reaction rate equation, the current value $q$
should be identified with $q^{\prime}$ not $q^{\prime\prime}$, since the
contrary treatment may lead to contradictions. For example, if $q=0$ at $t=0$
and $q$ is identified with $q^{\prime\prime}$ then the first reaction must be
emission with $q^{\prime\prime}=q^{\prime}+1$. We can set $q^{\prime}=0$ but
cannot assume $q^{\prime\prime}=0$ since $q^{\prime}=q^{\prime\prime
}-1=q-1=-1<0$ at $t=0$, which is unphysical. With $q^{\prime}=q,$ the
forward-time kinetics rate equation takes the form
\begin{align}
\frac{dq}{dt}  &  =k(q+1)f_{\text{A}^{\ast}}H\left(  f_{\text{A}^{\circ}%
}\right)  -kqf_{\text{A}^{\circ}}H\left(  f_{\text{A}^{\ast}}\right)
\nonumber\\
&  =k(q+1)f_{\text{A}^{\ast}}-kqf_{\text{A}^{\circ}}=k\left(  f_{\text{A}%
^{\ast}}-q\left(  f_{\text{A}^{\circ}}-f_{\text{A}^{\ast}}\right)  \right)
\text{ } \label{rmODE}%
\end{align}
where $k=K/\tau$ and the Heaviside constraints are not needed here since
equation (\ref{rmODE}) is formulated for solving it forward in time. Equation
(\ref{rmODE}) coincides with the Einstein theory of radiation
\cite{Einstein1917}. The same result has been obtained by using the quantum
field theory \cite{LL4} (due to pioneering work of Dirac \cite{Dirac1927}),
assuming coherent radiation in a Fock state $\left\vert q\right\rangle $ and
decohering matter. The photon creation and annihilation operators $\hat
{a}^{\dag}\left\vert q\right\rangle =(q+1)^{1/2}\left\vert q+1\right\rangle $
and $\hat{a}\left\vert q\right\rangle =(q)^{1/2}\left\vert q-1\right\rangle $
have the probability amplitudes consistent with the creation and annihilation
rates specified by (\ref{rmODE}). The quantum of energy associated with the
photon is transferred into decohered state of matter during absorption and
recoheres back into the Fock state during emission --- this explanation
matches the statistical interpretation given above, although decoherence and
recoherence are not explicitly considered or discussed in the standard
framework of second quantisation.

Assuming Gibbs distribution (\ref{Gibbs1}), where $\Delta E=h\nu,$ the
equilibrium solution is correctly given by the Bose-Einstein statistics
\cite{LL5}%
\begin{equation}
q^{(e)}=\frac{1}{\exp\left(  \frac{h\nu}{T}\right)  -1}\label{rmeq}%
\end{equation}
If the overall energy is constant, this distribution maximises the entropy
$S_{\nu}+S_{\text{A}},$ where
\begin{equation}
S_{\nu}=\left(  1+q\right)  \ln\left(  1+q\right)  -q\ln(q)\label{Snu}%
\end{equation}
and $S_{\text{A}}$ is defined by (\ref{Sx}). If many photon quantum states are
present $i=1,2,...,N_{q}$ then equation (\ref{rmODE}) needs to be written for
population $q_{(i)}$ at every state
\begin{equation}
\frac{dq_{(i)}}{dt}=k(q_{(i)}+1)f_{\text{A}^{\ast}}-kq_{(i)}f_{\text{A}%
^{\circ}},\text{ \ \ }i=1,...,N_{q}\label{rmODEn}%
\end{equation}
Evaluating the sum $Q=\Sigma_{i}q_{(i)}$ results in the common form of the
Einstein equation
\begin{equation}
\frac{dQ}{dt}=k(Q+N_{q})f_{\text{A}^{\ast}}-kQf_{\text{A}^{\circ}}\text{ }%
\end{equation}
Note that the stable equilibrium solution (\ref{rmeq}) exists only for
positive temperatures $T$. For negative temperatures $T<0$ in (\ref{Gibbs1}),
which corresponds to $f_{\text{A}^{\circ}}<f_{\text{A}^{\ast}}$ in
(\ref{rmODE}), equations (\ref{rmODEn}) are unstable:\ the photon populations
$q_{(i)}$ grow exponentially while being dominated by the few largest modes.
The growth is terminated only when the inverse population corresponding to
negative temperatures is exhausted, $f_{\text{A}^{\circ}}$ becomes larger than
$f_{\text{A}^{\ast}}$ and, then, a stable equilibrium can be achieved.
Practically, instability and exponential growth of major components are
observed in lasers.

The analysis of this section leads us to important conclusion

\begin{proposition}
Radiation in its interactions with matter (and presumably antimatter) is
decoherence-neutral: neither decoherence nor recoherence of radiation can
dominate unconditionally --- photons tend to decohere before and recohere
after the interaction events. Decohering neutrality of radiation matches the
Einstein theory of radiation and existing experimental evidence.
\end{proposition}

\section{Interactions of radiation and antimatter \label{Sra}.}

This section considers the interaction of radiation and matter assuming that
1) radiation is decoherence-neutral and 2) antimatter is dominated by
recoherence, in accordance with antisymmetric thermodynamics. The utility of
these assumptions is self-evident: under symmetric assumptions, interactions
of radiation and antimatter would simply be identical to the corresponding
interactions of radiation and matter.

The photon/antimatter interaction, which is shown in Figure \ref{fig5}
(bottom) is similar to the photon/matter interaction but antimatter recoheres
instead of decohering. Interactions of a single quantum state populated by $q$
photons with $F_{\text{\={A}}}$ antimatter atoms is specified by the
reactions
\begin{equation}
\text{1) \={A}}^{\circ}+\nu\longrightarrow\text{\={A}}^{\ast}\text{ \ \ and
\ \ 2) \={A}}^{\ast}\longrightarrow\text{\={A}}^{\circ}+\nu
\end{equation}

In this case, the number of reactions per reaction time $\tau$ is given by
\begin{equation}
\Delta q_{1}=Kf_{\text{\={A}}^{\ast}}^{\prime\prime}q^{\prime}H\left(
f_{\text{\={A}}^{\circ}}^{\prime}\right)  \label{ra1}%
\end{equation}
for the absorption reaction, where $f_{\text{\={A}}^{\ast}}^{\prime\prime
}\approx f_{\text{\={A}}^{\ast}}^{\prime}$ and $q^{\prime}=q^{\prime\prime
}+1\neq q^{\prime\prime}$ and
\begin{equation}
\Delta q_{2}=Kf_{\text{\={A}}^{\circ}}^{\prime\prime}q^{\prime\prime}H\left(
f_{\text{\={A}}^{\ast}}^{\prime}\right)  \label{ra2}%
\end{equation}
for the emission reaction, where $f_{\text{\={A}}^{\circ}}^{\prime\prime
}\approx f_{\text{\={A}}^{\circ}}^{\prime}$ and $q^{\prime\prime}=q^{\prime
}+1\neq q^{\prime}$. With $q^{\prime}=q$ and $\Delta q=\Delta q_{2}-\Delta
q_{1}$, the forward-time kinetics rate equation takes the form
\begin{align}
\frac{dq}{dt}  &  =k(q+1)f_{\text{\={A}}^{\circ}}H\left(  f_{\text{\={A}}%
^{\ast}}\right)  -kqf_{\text{\={A}}^{\ast}}H\left(  f_{\text{\={A}}^{\circ}%
}\right) \nonumber\\
&  \underset{f_{\text{\={A}}^{\circ}},f_{\text{\={A}}^{\ast}}>0}{=}k\left(
f_{\text{\={A}}^{\circ}}-q\left(  f_{\text{\={A}}^{\ast}}-f_{\text{\={A}}%
^{\circ}}\right)  \right)  \label{raODE}%
\end{align}
The Heaviside constraints cannot generally be omitted in this case.

A stable equilibrium solution of (\ref{raODE}) is possible only when
$f_{\text{\={A}}^{\ast}}>f_{\text{\={A}}^{\circ}}$ and $T=-\bar{T}>0$ in the
Gibbs distribution of antimatter%
\begin{equation}
\frac{f_{\text{\={A}}^{\ast}}}{f_{\text{\={A}}^{\circ}}}=\exp\left(
-\frac{\Delta E}{\bar{T}}\right)  \label{raeqf}%
\end{equation}
The solution is consistent and is specified by Bose-Einstein distribution
(\ref{rmeq}). If, however, $T=-\bar{T}<0$ and $f_{\text{\={A}}^{\ast}%
}<f_{\text{\={A}}^{\circ}},$ then equation (\ref{raODE}) is unstable and $q$
would grow exponentially similar to the exponential growth mentioned in the
previous section, although unlike in the previous section the process never
reaches a thermal equilibrium. As $q$ increases, $f_{\text{\={A}}^{\ast}}$
decreases and $f_{\text{\={A}}^{\circ}}$ increases\ due to conservation of
energy and this further stimulates emission specified by equation
(\ref{raODE}). Emission is terminated only when the excited population of
atoms is exhausted $f_{\text{\={A}}^{\ast}}=0$. At this stage, the antimatter
falls into the ground state. This behaviour is expected since the apparent
temperature of antimatter $T$ is a negative of its intrinsic temperature
$\bar{T}$ and, as noted in the previous section, the equilibrium with
radiation can be achieved only at positive apparent temperatures $T$.

The most interesting feature is that the step by step equations considered
here are exactly CPT-invariant: i.e. (\ref{rm1}) and (\ref{rm2}) can be
converted into (\ref{ra1}) and (\ref{ra2}) by swapping A with \={A} and $t$
with $-t$ (changing the direction of time implies swapping "prime" with
"double prime" and the forward reaction rate $\Delta q_{1}$ with the reverse
reaction rate $\Delta q_{2}$). Equations (\ref{rmODE}) and (\ref{raODE}),
however, do not represent the exact time inverse of each other and this is not
a mistake. Reaction rate equations (\ref{rmODE}) and (\ref{raODE}) are
formulated for conditions of dominant forward causality, when we set initial
conditions and not the final conditions. Both of these equations are to be
solved forward in time $t$. Equation (\ref{raODE}) would become the exact time
inverse of (\ref{rmODE}), if the antimatter equation was formulated for a
problem with final conditions. The environment, which has predominant forward
direction of thermodynamic time, interferes with our localised consideration
forcing us to set initial and not final conditions in our gedanken
experiments. As discussed in ref \cite{K-PhysA}, this kind of interference is
CP-invariant (and not CPT-invariant), which can lead to apparent CPT
violations: inconsistency between (\ref{rmODE}) and (\ref{raODE}) in the
present case. This violation is not a genuine violation of CPT symmetry --- it
would disappear if matter was changed to antimatter not only in our experiment
but also everywhere in the Universe \cite{K-PhysA}. The environmental
interference with the initial/final conditions, however, does not affect the
exact CPT symmetry of equation (\ref{sk}) derived for weak interactions of
matter and antimatter. Being decoherence-neutral, radiation is affected more
than matter and antimatter by the prevailing direction of decoherence in the environment.

The main outcomes of this section are summarised in form of the following proposition:

\begin{proposition}
The antisymmetric (CPT-invariant) approach to thermodynamics and kinetics
allows for stable \ equilibrium in interactions of radiation and antimatter
but only if the intrinsic temperature of antimatter is negative. Positive
intrinsic temperatures of antimatter result in radiation instabilities
bringing the antimatter into its intrinsic ground state.
\end{proposition}

Assuming that the overall energy is preserved, $F_{\text{\={A}}^{\ast}%
}+q=\operatorname{const}$ and that\ $F_{\text{\={A}}^{\ast}},dF_{\text{\={A}}%
^{\circ}}>0$, we can write%
\begin{equation}
-\frac{dF_{\text{\={A}}^{\ast}}}{dt}=\frac{dF_{\text{\={A}}^{\circ}}}%
{dt}=\frac{dq}{dt}=k\left(  (q+1)f_{\text{\={A}}^{\circ}}-qf_{\text{\={A}}%
^{\ast}}\right)
\end{equation}
considering intrinsic entropies $S_{\nu}$ and $S_{\operatorname{\bar{A}}}$
defined by (\ref{Snu}) and (\ref{Sx}), the H-theorem can be formulated as

\begin{proposition}
For interactions of radiation and antimatter, the symmetric $S_{s}=S_{\nu
}+S_{\operatorname{\bar{A}}}$ and antisymmetric $S_{a}=S_{\nu}%
-S_{\operatorname{\bar{A}}}$ entropies are increased forward in time $t$ by,
correspondingly, symmetric and antisymmetric kinetics until the evolution
reaches its equilibrium or is terminated by physical constraints.
\end{proposition}

Since the case of symmetric kinetics is equivalent to interactions of
radiation and matter and, thus, is obvious, we prove the H-theorem only for
the antisymmetric case:
\begin{align}
\frac{dS_{a}}{dt}  &  =\frac{dq}{dt}\ln\left(  \frac{1+q}{q}\right)
+\sum_{\text{X=\={A}}^{\ast},\text{\={A}}^{\circ}}\frac{dF_{_{\text{X}}}}%
{dt}\left(  \ln\left(  f_{\text{X}}\right)  +1\right)  =\nonumber\\
&  =k\left(  (q+1)f_{\text{\={A}}^{\circ}}-qf_{\text{\={A}}^{\ast}}\right)
\ln\left(  \frac{\left(  1+q\right)  f_{\text{\={A}}^{\circ}}}%
{qf_{\text{\={A}}^{\ast}}}\right)  \geq0
\end{align}
for any positive $q,$ $f_{\text{\={A}}^{\ast}}$ and $f_{\text{\={A}}^{\circ}}%
$. The evolution is terminated $dS_{a}/dt=0$ if the physical boundary (i.e.
$F_{\text{\={A}}^{\ast}}=0$ or $F_{\text{\={A}}^{\circ}}=0$) is reached.

\section{Discussion and conclusions}

One of the main conclusions of the present work in the context of interactions
of radiation and matter is neutrality of radiation with respect to decoherence
and recoherence: neither of these processes can dominate the other. Only
decoherence-neutral treatment of radiation matches the Einstein theory
\cite{Einstein1917} and experimental evidence \cite{light1981}. While our
world is dominated by decoherence, presence of recoherence in interactions of
radiation with matter was envisaged a hundred years ago by the insights of the
Einstein radiation theory and later experimentally confirmed by existence of
lasers. Unlike low-temperature Bose condensates, which are forced into a
single quantum state by the energy constraint, the most spectacular display of
dynamic recoherence in lasers occurs at high temperatures with plenty of
quantum states to select from. Does the conclusion of decohering neutrality of
radiation have any implications for choosing between symmetric and
antisymmetric extensions of thermodynamics into antimatter?

\begin{figure}[h]
\begin{center}
\includegraphics[width=7cm,page=5,trim=4cm 4cm 4cm 1cm, clip ]{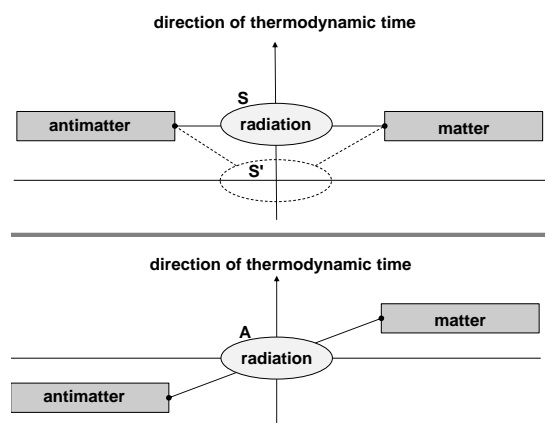}
\caption{Directions of thermodynamic time for matter, antimatter and radiation. Top: symmetric (CP-invariant) approach; bottom: antisymmetric (CPT-invariant) approach. }
\label{fig6}
\end{center}
\end{figure}

According to the antisymmetric view of thermodynamic interactions of matter
and antimatter, decoherence occurs in opposite temporal directions for matter
and antimatter. Since radiation is neutral with respect to the
matter/antimatter duality, there could not be a direction of decoherence
intrinsically associated with radiation (Figure \ref{fig6}, bottom). Yet,
radiation must decohere during interactions with matter (antimatter); at least
to some extent since, otherwise, concepts of statistical physics would not be
applicable to these interactions, which is obviously incorrect. In the
antisymmetric treatment of the problem, radiation does not have any intrinsic
decohering mechanisms on its own, but interactions of radiation with matter
(antimatter) should cause radiation decoherence and recoherence.

While decohering neutrality of radiation and physical existence of recoherence
make this antisymmetric approach to thermodynamics and kinetics rather
attractive, it does not prove the approach. Matter and antimatter can still
have the same direction of the thermodynamic time, while radiation having no
temporal preference as shown by in Figure \ref{fig6} (top, case S$^{\prime}$).
The neutrality of radiation only forces us to discard theories presuming that
decoherence is caused by interactions with universal fields (such as gravity
-- see ref.\cite{PenroseBook}) since photons should not interact with these
fields. However, other fields (e.g. the Higgs field) may have the desired
decohering effect on particles and antiparticles, but no effect on photons.

While the symmetric and environmental mechanisms of decoherence are more
conceptually aligned with presence of a time priming field, and the
antisymmetric decoherence generally points in the direction of intrinsic
mechanisms as possible sources of the time asymmetry, this work concludes that
both mechanisms are likely to be important. The major challenge in
experimental testing of the invariant properties of thermodynamics is in
creating an antisystem --- a thermodynamic system made of antimatter. Since
the mechanism of thermodynamic time remains unknown, it is not clear how large
this antisystem needs to be. One of the smallest possible antisystems might be
represented by quark-gluon plasma produced in collision of antibaryons,
although the extent of thermodynamic properties that can exists at these very
small scales is not fully known \cite{Nature2007}.

It seems that the best approach for testing the invariant properties of
thermodynamics is in trapping a statistical set of antiatoms, investigating
their emission/absorption properties and comparing them with the conventional
thermodynamic properties for emission/absorption by a similar set of atoms.
The analysis of Sections \ref{Srm} and \ref{Sra} indicates that, in symmetric
thermodynamics, both sets should have exactly the same properties. However,
the antiatoms are expected to fall into their intrinsic ground states
according to antisymmetric thermodynamics. This does not mean that a
particular antiatom cannot be in an excited state but only that this state is
statistically unlikely in thermodynamic antisystems. The absorption/emission
properties of any isolated quantum objects and antiobjects, which do not
display thermodynamic properties, are expected to be CP-invariant; it is only
thermodynamic antisystems that behave unconventionally in antisymmetric
thermodynamics. Since the antiatoms are likely to be gathered in very small
quantities, reaching the statistical scales becomes a major issue.
Antisymmetric thermodynamics predicts that an antiatom system would be more
reluctant to absorb light than the corresponding atom system, and this
asymmetry should be evident for systems that are sufficiently large to exhibit
thermodynamic properties. In antisymmetric thermodynamics, equilibrium between
radiation and antimatter is theoretically possible but (under normal
conditions) not practically, since this equilibrium requires negative
intrinsic temperatures of antimatter. Rapid progress in antimatter creation
and spectroscopy suggests that it might be possible to conduct these
experiments with existing levels of technology \cite{H2anti2010,MuB2011}.

\section*{Acknowledgement}

The author thanks Dimitri A. Klimenko for numerous discussions. The author
also appreciates constructive comments of the anonymous reviewers.

\bibliographystyle{mdpi}
\bibliography{Law3}
\bigskip

\appendix

\section*{Appendix: Quantum typicality, mixtures and decoherence \label{A1}}

Canonical typicality is a family of important theorems and results that have
been independently introduced by a number of authors
\cite{Lloyd1988,CT-G2006,CT-P2006} and can relate typical quantum states of
large systems to thermodynamic properties. Only microcanonical distributions,
which correspond to maximally mixed state (a quantum mixture with the same
probability of all quantum states) are considered here, although typicality
can be also extended to other canonical distributions \cite{CT-G2006}. We
discuss these issues with a minor alteration: as explained below, canonical
typicality is meaningful not only for pure states but also for mixtures.

Considering a Hilbert space $\mathcal{H}_{0}$ of large dimension $n_{0}$, a
mixture of $n_{m}$ wave function $\psi^{l}$ the corresponding density matrix
$\boldsymbol{\rho}$\ and Hermitian operator $\mathbb{G}$ with eigenvalues
$g_{i}$ and eigenstates $\left\vert g_{i}\right\rangle $ measuring $G$
\begin{equation}
G=\sum_{l=1}^{n_{m}}\left\langle \psi^{l}|\mathbb{G}|\psi^{l}\right\rangle
=\sum_{l=1}^{n_{m}}\sum_{i=1}^{n_{0}}\left\vert \left\langle g_{i}|\psi
^{l}\right\rangle \right\vert ^{2}g_{i}=\text{Tr}\left[  \mathbb{G}%
\boldsymbol{\rho}\right]  ,\text{ }\boldsymbol{\rho}=\sum_{l=1}^{n_{m}%
}\left\vert \psi^{l}\right\rangle \left\langle \psi^{l}\right\vert
\end{equation}
the following relations are valid for any $\mathbb{G}$
\begin{equation}
G_{0}=\underset{\boldsymbol{\operatorname{Tr}[\rho]}=1}{\text{Mean}}\left[
\sum_{l=1}^{n_{m}}\left\langle \psi^{l}|\mathbb{G}|\psi^{l}\right\rangle
\right]  =\frac{\text{Tr}\left[  \mathbb{G}\right]  }{n_{0}} \label{G1}%
\end{equation}
and
\begin{equation}
\Delta G_{0}=\underset{\boldsymbol{\operatorname{Tr}[\rho]}=1}{\text{RMS}%
}\left[  \sum_{l=1}^{n_{m}}\left\langle \psi^{l}|\mathbb{G}|\psi
^{l}\right\rangle -\frac{\text{Tr}\left[  \mathbb{G}\right]  }{n_{0}}\right]
\sim\frac{\text{Tr}\left[  \mathbb{G}^{2}\right]  ^{1/2}}{n_{0}n_{m}^{1/2}}
\label{G2}%
\end{equation}
Here, $n_{m}$ is assumed to be fixed and the states are selected at random
provided they are compliant with normalisation condition
$\boldsymbol{\operatorname{Tr}[\rho]}=1,$ that is
\begin{equation}
\psi^{l}=\sum_{l=1}^{n_{m}}\sum_{i=1}^{n_{0}}c_{i}^{l}\left\vert
g_{i}\right\rangle ,\ \ \ \sum_{l=1}^{n_{m}}\sum_{i=1}^{n_{0}}\left\vert
c_{i}^{l}\right\vert ^{2}=1,\ \ \ G=\sum_{l=1}^{n_{m}}\left\langle \psi
^{l}|\mathbb{G}|\psi^{l}\right\rangle =\sum_{l=1}^{n_{m}}\sum_{i=1}^{n_{0}%
}g_{i}\left\vert c_{i}^{l}\right\vert ^{2}%
\end{equation}
Pure states of the system correspond to $n_{m}=1$.

The statements can be proved by using normalised eigenstates $\left\vert
g_{i}\right\rangle $ of the operator $\mathbb{G}$ and extending the space
under consideration. Let $\mathbb{\acute{G}}=\mathbb{G}\otimes\mathbb{I}_{m},$
having a set of orthogonal eigenvectors $\left\vert \acute{g_{i}}\right\rangle
=\left\vert g_{j}\right\rangle \otimes\left\vert l\right\rangle ,$
$j=1,...,n_{0},$ $l=1,...,n_{m},$ $i=1,...,n_{1}$, operate on the extended
Hilbert space $\mathcal{\acute{H}}=\mathcal{H}_{0}\otimes\mathcal{H}_{m}$ of
dimension $n_{1}=n_{0}n_{m},$ where $\mathcal{H}_{m}$ has the
basis$\ \left\vert l\right\rangle $ and the dimension $n_{m}$. The action of
the operator $\mathbb{G}$ can be equivalently represented by the following
equations applying $\mathbb{\acute{G}}$ to pure state $\acute{\psi}:$%
\begin{equation}
\acute{\psi}=\sum_{i=1}^{n_{1}}\acute{c}_{i}\left\vert g_{i}\right\rangle
,\ \ \ \sum_{i=1}^{n_{1}}\left\vert \acute{c}_{i}\right\vert ^{2}%
=1,\ \ \ G=\left\langle \acute{\psi}|\mathbb{\acute{G}}|\acute{\psi
}\right\rangle =\sum_{i=1}^{n_{1}}\acute{g}_{i}\left\vert \acute{c}%
_{i}\right\vert ^{2}%
\end{equation}
where $n_{1}=n_{0}n_{m}$ and $\acute{c}_{j(i,l)}=c_{i}^{l}$ are the same
coefficients.\ In the extended space, the equations (\ref{G1}) and (\ref{G2})
become
\begin{equation}
G_{0}=\underset{\left\Vert \acute{\psi}\right\Vert =1}{\text{Mean}}\left[
\left\langle \acute{\psi}|\mathbb{\acute{G}}|\acute{\psi}\right\rangle
\right]  =\sum_{i=1}^{n_{1}}\acute{g}_{i}\underset{\left\Vert \acute{\psi
}\right\Vert =1}{\text{Mean}}\left[  \left\vert \acute{c}_{i}\right\vert
^{2}\right]  =\sum_{i=1}^{n_{1}}\frac{\acute{g}_{i}}{n_{1}}=\frac
{\text{Tr}\left[  \mathbb{\acute{G}}\right]  }{n_{1}}=\frac{\text{Tr}\left[
\mathbb{G}\right]  }{n_{0}}%
\end{equation}%
\[
\Delta G_{0}=\underset{\left\Vert \acute{\psi}\right\Vert =1}{\text{Mean}%
}\left[  \left(  \left\langle \acute{\psi}|\mathbb{\acute{G}}|\acute{\psi
}\right\rangle -\frac{\text{Tr}\left[  \mathbb{\acute{G}}\right]  }{n_{1}%
}\right)  ^{2}\right]  =\underset{\left\Vert \acute{\psi}\right\Vert
=1}{\text{Mean}}\left[  \left\langle \acute{\psi}|\mathbb{\acute{G}}%
|\acute{\psi}\right\rangle ^{2}\right]  -\frac{\text{Tr}\left[  \mathbb{\acute
{G}}\right]  ^{2}}{n_{1}^{2}}=
\]%
\begin{equation}
=\sum_{i=1}^{n_{1}}\sum_{j=1}^{n_{1}}\acute{g}_{i}\acute{g}_{j}%
\underset{\left\Vert \acute{\psi}\right\Vert =1}{\text{Mean}}\left[
\left\vert \acute{c}_{i}\right\vert ^{2}\left\vert \acute{c}_{j}\right\vert
^{2}-\frac{1}{n_{1}}\frac{1}{n_{1}}\right]  \sim\frac{\text{Tr}\left[
\mathbb{\acute{G}}^{2}\right]  }{n_{1}^{2}}=\frac{\text{Tr}\left[
\mathbb{G}^{2}\right]  }{n_{0}^{2}n_{m}}%
\end{equation}
since the last double sum is dominated at the limit $n_{1}\rightarrow\infty
$\ by the diagonal $j=i$ elements involving $\left[  \left\vert \acute{c}%
_{i}\right\vert ^{4}-1/n_{1}^{2}\right]  $, while Tr$\left[  \mathbb{I}%
_{m}\right]  =n_{m}$. More accurate estimates of the variance and more
detailed derivations can be found in Lloyd's PhD thesis \cite{Lloyd1988}.

In the case of pure states $n_{m}=1$, the mean is taken over all possible wave
functions $\psi$ that satisfy the normalisation constraint $\left\Vert
\psi\right\Vert =1$. For many (but not for all) operators $\mathbb{G}$ the
average $G=G_{0}$ provides a very good estimate for the measurement $G(\psi)$
when $\psi$ is fixed (but may be unknown) since $\Delta G_{0}\ll G_{0}$.\ This
requires that Tr$\left[  \mathbb{G}^{2}\right]  ^{1/2}\ll$Tr$\left[
\mathbb{G}\right]  $; that is, the measurement operation must involve a large
number of $g_{i}\sim g>0$ that are not dominated by few of these values.
Canonical typicality is a quantum version of the laws of large numbers,
reflecting summation over many uncertain parameters. For most functions
normalised by $\left\Vert \psi\right\Vert =1,$ the value of $G(\psi)$ is close
to its average, we will call such functions typical and the remaining
functions atypical.

Typicality of $\psi$ generally depends on the operator $\mathbb{G}$ under
consideration, and canonical typicality does not work for all possible
$\mathbb{G}$. For example, $g_{1}=1$ and $g_{2}=g_{3}=...=0$ would produce for
$n_{m}=1$ the mean of $G_{0}=g_{1}/n_{0}$ with $\Delta G_{0}\sim G_{0}$.
However, if the system is in a genuine maximally mixed state (which
corresponds to a mixture of $n_{m}=n_{0}$ wave functions with
$\boldsymbol{\rho}=\mathbf{I}/n_{0}$), then $G_{0}=g_{1}/n_{0}$ and $\Delta
G_{0}=0$. A similar result $G_{0}=g_{1}/n_{0}$ and $\Delta G_{0}\sim
G_{0}/n_{m}^{1/2}\approx0$ can be obtained when the system is a mixture of a
substantial number $n_{m}$ (which, however, can be incomparably smaller than
$n_{0}$) of typical pure states. This illustrates that, for a very large
system, distinguishing a typical pure state from the maximally mixed state by
performing measurements is difficult but conceptually possible. A mixture
dominated by typical states, which nevertheless might be very far from the
maximally mixed state $1\ll n_{m}\ll n_{0}$, cannot practically be
distinguished from the maximally mixed state by measuring observables.

Another issue is that every particular quantum measurement over a pure state
$\psi$ produces not $G=\left\langle \psi|\mathbb{G}|\psi\right\rangle $\ but
$g_{i}$ with probability $p_{i}=\left\vert \left\langle g_{i}|\psi
\right\rangle \right\vert ^{2}$. Hence, to determine $G$ for given $\psi,$ we
need to set many realisations of the experiment with exactly the same initial
wave function $\psi,$ which is often practically impossible for large systems.
Selecting various unknown $\psi$ for initial conditions has the same effect as
having a quantum mixture of these functions. When performing quantum
measurements of a large system, distinguishing the maximally mixed state from
typical pure wave functions or their mixtures is practically difficult.

Canonical typicality becomes most useful when we are interested only in
characteristics of subsystems that are much smaller than their very large
parent systems. In this case, the Hilbert space of the overall system
$\mathcal{H}_{0}$ be represented $\mathcal{H}_{0}=\mathcal{H}_{S}%
\otimes\mathcal{H}_{E}$ as tensor product of Hilbert spaces of a subsystem and
its environment. The dimension of the environment space is $n_{E}=n_{0}/n_{S}$
where $n_{0}$ and $n_{S}$\ are dimensions of $\mathcal{H}_{0}$ and
$\mathcal{H}_{S}$. It is easy to see that any operator $\mathbb{G}_{S}$ acting
on subspace $\mathcal{H}_{S}$ (which is necessarily extended to the whole
space $\mathbb{G=G}_{S}\otimes\mathbb{I}_{E}$ by the corresponding unit
operator $\mathbb{I}_{E}$) involves summation of a large number of states from
$\mathcal{H}_{E}$ and is compliant with the condition Tr$\left[
\mathbb{G}\right]  \gg$Tr$\left[  \mathbb{G}^{2}\right]  ^{1/2}$\ for
$G_{0}\gg\Delta G_{0}$. The real and imaginary parts of elements $\rho
_{S}^{kj}$ of reduced density matrix $\boldsymbol{\rho}_{S}=$Tr$_{\mathcal{H}%
_{E}}\left[  \boldsymbol{\rho}\right]  $ can be evaluated with assistance of
the Hermitian operators $2\mathbb{G}_{+}^{kj}=\left(  \left\vert
k\right\rangle \left\langle j\right\vert +\left\vert j\right\rangle
\left\langle k\right\vert \right)  \otimes\mathbb{I}_{E}$ and $2\mathbb{G}%
_{-}^{kj}=i\left(  \left\vert k\right\rangle \left\langle j\right\vert
-\left\vert j\right\rangle \left\langle k\right\vert \right)  \otimes
\mathbb{I}_{E}$ (essentially, $\rho_{S}^{kj}=$Tr$\left[  \mathbb{G}_{+}%
^{kj}\boldsymbol{\rho}\right]  +i$Tr$\left[  \mathbb{G}_{-}^{kj}%
\boldsymbol{\rho}\right]  $).\ Since Tr$\left[  \mathbb{G}_{+}^{jj}\right]
=n_{E},$ Tr$\left[  \mathbb{G}_{+}^{kj}\right]  =0$ ($k\neq j$), Tr$\left[
\mathbb{G}_{-}^{kj}\right]  =0$ and Tr$\left[  \mathbb{G}_{\pm}^{kj}%
\mathbb{G}_{\pm}^{kj}\right]  =n_{E},$\ we obtain from (\ref{G1}) and
(\ref{G2})%
\begin{equation}
\text{Mean}\left[  \rho_{S}^{kj}\right]  =\frac{\delta^{kj}}{n_{S}%
},\ \ \ \text{RMS}\left[  \rho_{S}^{kj}-\frac{\delta^{kj}}{n_{S}}\right]
\sim\frac{1}{n_{S}n_{E}^{1/2}n_{m}^{1/2}} \label{rhoS}%
\end{equation}
where $\delta^{kj}$ is Kronecker delta denoting components of $\mathbf{I}%
_{S}.$ This indicates mean-square convergence of $\boldsymbol{\rho}_{S}$ to
$\mathbf{I}_{S}/n_{S}$ as $n_{E}\rightarrow\infty$. For more rigorous
estimates and convergence in probability see ref.\cite{CT-P2006}.

In a multi-particle system, a pair of two particles can be thought to be a
subsystem. If the pair has two particles A and B then $\mathcal{H}%
_{S}=\mathcal{H}_{\text{A}}\otimes\mathcal{H}_{\text{B}}.$ In the case
specified by (\ref{rhoS}), the reduced density matrix can be factorised
$\boldsymbol{\rho}_{S}=\boldsymbol{\rho}_{\text{AB}}=\boldsymbol{\rho
}_{\text{A}}\otimes\boldsymbol{\rho}_{\text{B}}$. Hence, the states of
particles A and B are separable and particles behave in a statistically
independent manner. It is worthwhile to note that the selected pair of
particles A and B is, effectively, in the maximally mixed state while the
overall state of the system can be both mixed or pure. What is important that
the pure state is typical (such as selected at random) or that the mixture is
dominated by the typical states.

The assumption of canonical typicality indicates that the tested subsystem
displays thermodynamic properties. Our consideration, however, is based on
explicit use of decoherence, which necessarily converts pure quantum states
into true mixtures. Canonical typicality establishes that typical pure states
and, especially, mixtures of typical states are difficult to distinguish from
the maximally mixed state. Consideration of decoherence is needed not because
a measured property $\mathbb{G}$ of typical pure states and maximal mixtures
would be significantly different at a given time --- canonical typicality
states that this is rather unlikely --- but because time-directional
properties of decoherence and recoherence cause radically different temporal
evolutions of the systems. We note that canonical typicality does not
discriminate the directions of time --- typical states do not develop or
disappear but simply exist because they are more numerous and likely than
atypical states. If decoherence is not considered, the direction of
thermodynamic time must be introduced by other means. Discrimination of time
directions is the first necessary assumption for thermalisation, which is
commonly based on implicitly introduced discriminating measures such as
specifying initial and not final conditions. This may be practically effective
for many problems but it inevitably restricts our understanding of time by the
boundaries of common intuition based on causality. There are other constraints
imposed by unitary evolutions. First, a quantum system must eventually recur
to the proximity of its original state due to the quantum recurrence theorem
\cite{QRT1957}. Second, a quantum system in a maximally mixed state can be
evolved backward in time obtaining the obvious outcome: the system stays in
the maximally mixed state. Hence demonstration of thermalisation must involve
additional assumptions such as temporal averaging, coarsening of the solution
or some external interference that can produce effects similar to decoherence.

\begin{figure}[h]
\begin{center}
\includegraphics[width=10cm,page=6,trim=3cm 4cm 0cm 4.3cm, clip ]{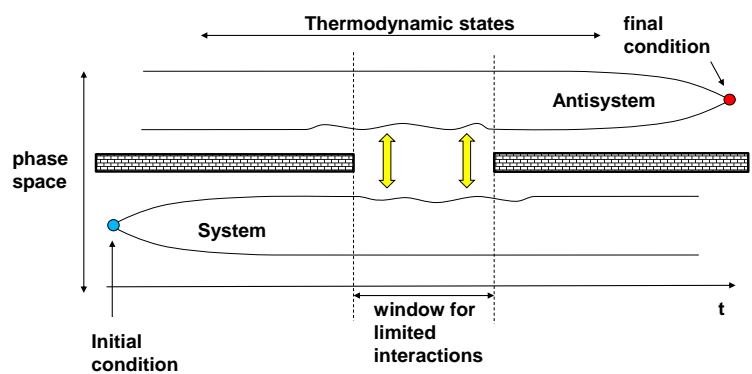}
\caption{Possible conditions for ensuring thermodynamic states of a system and an antisystem during indirect and limited interactions (which prevent annihilation).}
\label{fig7}
\end{center}
\end{figure}

The question if canonical or microcanonical statistical states can be achieved
when started from an atypical initial state (i.e. thermalisation) is an
important question, which has been repeatedly discussed in the literature
\cite{CT-P2009,Yukalov2012,QantumThermo2016}. This question generally remains
outside the scope of this work: closeness to the maximally mixed state was
assumed in Section 2 (on the basis of the understanding of canonical
typicality presented here). We, however, note here that decoherence is likely
to play a significant role in thermalisation. Indeed, maximally mixed states
are asymptotically achieved in evolutions predicted by the symmetric Pauli
master equation, when the decoherence basis does not completely coincide with
the energy basis, that is when there exist some energy interactions between
persistently decohering components. Under these conditions, the Pauli master
equation unambiguously predicts convergence to a maximally mixed state
(without any need for time averaging and other assumptions)
\cite{Pauli1928,SciRep2016}. In the context of antisymmetric thermodynamics,
considered here, the situation is more complicated. The thermodynamic states
of the system and antisystem can be ensured (at least theoretically -- see
Figure \ref{fig7}) by setting initial conditions for the system and the final
conditions for the antisystem and allowing limited interactions between the
system and antisystem within a time window well in between the initial and
final moments.

While there are a few plausible assumptions that, at least in principle, can
be used to explain the workings of thermodynamic time, sooner or later modern
physics will achieve a stage when these assumptions can be tested
experimentally. While it is obvious that thermodynamic time "flows" (i.e.
increases entropy) from the past to future, even the order of magnitude of the
process priming this flow remains unknown --- it is difficult to search for a
needle in a haystack when the parameters of that needle are not known. We are
aware of two types of experiments that might (just might) bring new evidence
into this matter. First, if detected, any asymmetry of thermodynamic
properties of matter and antimatter would be a reflection of an intrinsic
time-priming mechanism. Second, time-priming environmental interference may be
detected in CP-violating and CPT-preserving quantum processes as apparent CPT
violations \cite{K-PhysA}. While a number of CP violations have been detected
in K- and B- mesons \cite{PDG2012}, the most recent examination of mixing in
neutral B-mesons \cite{Barbar2016}, indicates presence of a larger than
previously thought uncertainty in a CPT-violating parameter.

\end{document}